\documentstyle[prl,floats,aps,twocolumn,epsf,graphicx]{revtex}
\begin{document}
\twocolumn[\hsize\textwidth\columnwidth\hsize\csname
@twocolumnfalse\endcsname

\title{Temporal oscillations and phase transitions in the evolutionary minority game}
\author{Ehud Nakar$^1$ and Shahar Hod$^2$}
\address{$^1$The Racah Institute of Physics, The
Hebrew University, Jerusalem 91904, Israel}
\address{}
\address{$^2$Department of Condensed Matter Physics, Weizmann Institute, Rehovot
 76100, Israel}
\date{\today}
\maketitle

\begin{abstract}

\ \ \ The study of societies of adaptive agents seeking minority status is an active
area of research. Recently, it has been demonstrated that such systems display 
an intriguing phase-transition: agents tend to {\it self-segregate}
or to {\it cluster} according to the value of the prize-to-fine ratio, $R$. 
We show that such systems do {\it not} establish a true stationary distribution. 
The winning-probabilities of the agents display temporal oscillations. The amplitude and frequency of the oscillations 
depend on the value of $R$. The temporal oscillations which characterize the system explain 
the transition in the global behavior from self-segregation to clustering in the $R<1$ case.

\end{abstract}
\bigskip

]

\section{Introduction}

The study of complex systems is a growing area of research. A
problem of central importance in biological and socio-economic systems
is that of an evolving population in which individual agents adapt their
behavior according to past experience. 
Of particular interest are situations in which members (usually referred to
as `agents') compete for a limited resource, or to be in a minority (see e.g.,
\cite{John1} and references therein.) In financial markets for instance, more
buyers than sellers implies higher prices, and it is therefore better for a
trader to be in a minority group of sellers. Predators foraging for food will
do better if they hunt in areas with fewer competitors. Rush-hour drivers, facing
the choice between two alternative routes, wish to choose the route containing
the minority of traffic \cite{HubLuk}. 

Considerable progress in the theoretical understanding of such systems has been
gained by studying the simple, yet realistic model of the minority game (MG) 
\cite{ChaZha}, and its evolutionary version (EMG) \cite{John1} 
(see also \cite{DhRo,BurCev,LoHuJo,HuLoJo,HaJeJoHu,BuCePe,LoLiHuJo,LiVaSa,SaMaRi}
and references therein). The EMG consists of an odd number of \( N $ agents
repeatedly choosing whether to be in room ``0'' (e.g., choosing to sell an asset
or taking route A) or in room ``1'' (e.g., choosing to buy an asset or taking
route B). At the end of each round, agents belonging to the smaller group
(the minority) are the winners, each of them gains $R$ points (the ``prize''),
while agents belonging to the majority room lose $1$ point (the ``fine'').
The agents have a common ``memory'' look-up table, containing the outcomes of
$m$ recent occurrences. Faced with a given bit string of recent $m$
occurrences, each agent chooses the outcome in the memory with probability $p$,
known as the agent's ``gene'' value (and the opposite alternative with
probability $1-p$). If an agent score falls below some value $d$,
then its strategy (i.e., its gene value) is modified (One can also speaks in terms 
of an agent quitting the game, allowing a new agent to take his place.) 
In other words, each agent tries to learn from his past mistakes, and
to adjust his strategy in order to survive. 

Early studies of the EMG were restricted to simple situations in which the prize-to-fine 
ratio $R$ was assumed to be equal unity (see however \cite{BurCev}). 
A remarkable conclusion deduced from the EMG \cite{John1} 
is that a population of competing agents tends to {\it self-segregate} into opposing 
groups characterized by extreme behavior. 
It was realized that in order to flourish in such situations, an agent 
should behave in an extreme way ($p=0$ or $p=1$) \cite {John1,note1}. 

On the other hand, in many real life situations the prize-to-fine 
ratio may take a variety of different values \cite{HodNak}. A different kind of 
strategy may be more favorite in such situations. 
In fact, we know from real life situations that extreme behavior is not always optimal. 
In particular, our daily experience indicates that in difficult situations 
(e.g., when the prize-to-fine ratio is low) human people tend to be 
confused and indecisive. In such circumstances 
they usually seek to do the {\it same} (rather than the opposite) as the majority. 

Based on this qualitative expectation, we have recently extended the 
exploration of the EMG to generic situations in which the prize-to-fine ratio $R$ takes 
a variety of different values. 
It has been shown \cite{HodNak} that a sharp phase transition exist in the model: 
``confusion'' and ``indecisiveness'' take over in times of depression (for which 
the prize-to-fine ratio is smaller than some critical value $R_c$), in 
which case central agents (characterized by $p={1 \over 2}$) 
perform better than extreme ones. That is, for $R<R_c$ agents tend to 
{\it cluster} around $p={1 \over 2}$ (see Fig. 1 in \cite{HodNak}) rather than self-segregate 
into two opposing groups.

In this paper we provide an explanation for the global behavior of agents in 
the EMG. The model is 
based on the fact that the population never establishes a true stationary 
distribution. In fact, the probability of a particular agent to win, $\tau (p)$, 
is {\it time-dependent}. This fact has been overlooked in former studies of the EMG. 
The winning-probability oscillates in time: the amplitude and frequency of 
the oscillations depend on both the value of the prize-to-fine ratio $R$ and on 
the agent's gene value $p$. 
The smaller the value of $R$ the larger is the oscillation amplitude. 
In addition, ``extreme'' agents (with $p=0,1$) have an 
oscillation amplitude which is larger than the corresponding amplitude of ``central'' 
agents (those with $p={1 \over 2}$). 

We show that in the $R>R_c$ case these oscillations are used by extreme agents 
to cooperate indirectly and to share the system's resources efficiently. 
On the other hand, when $R<R_c$ agents cannot afford to share the 
limited resources. They tend to cluster around $p={1 \over 2}$, preventing any possibility of 
cooperation.

\section{Temporal oscillations of the winning-probabilities}

A partial explanation for the ({\it steady state}) gene-distribution is 
given in \cite{LoHuJo}. It has been found that the probability $\tau (p)$ 
of an agent with a gene-value $p$ to win is given by:

\begin{equation}\label{Eq1}
\tau(p)=1/2-\alpha p(1-p)\  ,
\end{equation}
where $\alpha <1$ is a constant (which depends on the number of agents $N$). 
This result is used to explain the better performance of extreme agents 
as compared to central ones, which leads to the 
phenomena of self-segregation \cite{LoHuJo}. 
However, the analytic model presented in \cite{LoHuJo} cannot explain the 
phase transition (from self-segregation to clustering) observed 
in the exact model \cite{HodNak}.

In Figure 4 of \cite{HodNak} we have displayed the time-dependence of the average gene value, $<$$p$$>$, 
for different values of the prize-to-fine ratio $R$. It has been demonstrated that 
the distribution $P(p)$ oscillates around $p={1 \over 2}$. 
The smaller the value of $R$, the larger are the amplitude and the 
frequency of the oscillations. 
Thus, we conclude that a population which evolves in a tough environment 
never establishes a steady state distribution. Agents are constantly changing 
their strategies, trying to survive. By doing so they create global currents in 
the gene space. 

\begin{figure}
\centerline{\epsfxsize=9cm \epsfbox{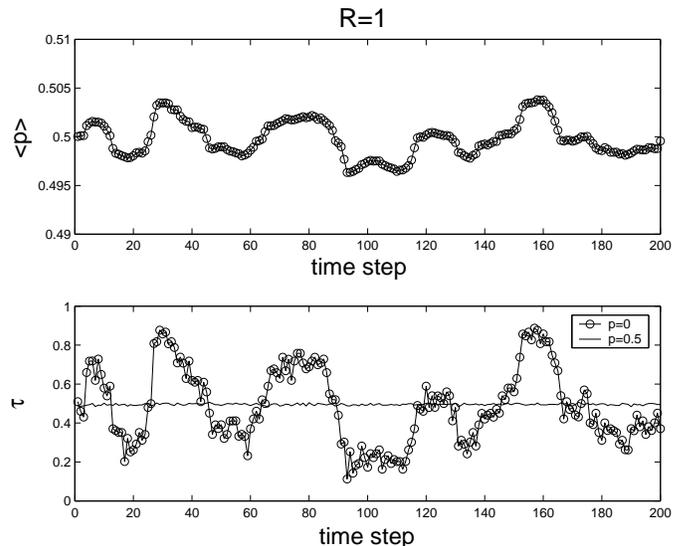}} 
\caption{Temporal dependence of the winning probabilities for $R=1$. 
The results are for $N=10001$ agents, and $d=-4$. 
$\tau (p=0)$ oscillates in time, with an amplitude of $\sim 0.3$, while 
$\tau (p={1 \over 2})$ is practically constant ($\sim 0.5$) in time. 
The period of the oscillations is about $40$ time steps.}
\label{Fig1}
\end{figure}

\begin{figure}
\centerline{\epsfxsize=9cm \epsfbox{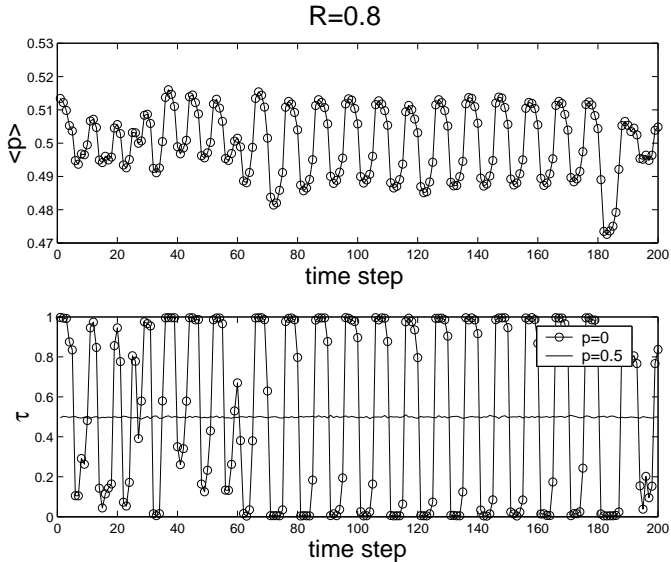}} 
\caption{Temporal dependence of the winning probabilities for $R=0.8$. 
The parameters are the same as in Fig. \ref{Fig1}.
$\tau (p=0)$ oscillates in time, with the maximally possible amplitude 
of $\sim 0.5$, while 
$\tau (p={1 \over 2})$ is practically constant ($\sim 0.5$) in time. 
The period of the oscillations is about $10$ time steps.}
\label{Fig2}
\end{figure}

The temporal oscillations of $<$$p$$>$ induce larger oscillations in the 
winning-probabilities $\tau (p)$ of the agents. In Fig. \ref{Fig1} we display the temporal 
dependence of $\tau (p=0)$ and $\tau (p={1 \over 2})$ for $R=1$. 
Figure \ref{Fig2} displays the same quantities for the case of $R=0.8$. Both figures are 
produced from exact numerical simulations of the the EMG. 
We find that when $<$$p$$>$ is even slightly higher than ${1 \over 2}$, $\tau (p=0)$ 
(the winning-probability of an agent who acts against the global memory outcome) 
is almost unity.

It should be emphasized that the winning probability of a central agent, $\tau({1 \over 2})$, 
displays only mild oscillations. For a central agent (characterized by $ p\simeq {1 \over 2}$) 
it is basically irrelevant which room is more probable to win in the next round of the game. 
In either case, his winning-probability is approximately ${1 \over 2}$. 
In other words, the global gene-distribution of the population has a larger influence on extreme
agents as compared to central ones.

It is evident from Figs. \ref{Fig1} and \ref{Fig2} 
that the amplitude and the frequency of the oscillations increase as
the value of the prize-to-fine ratio $R$, decreases.
It is important to note 
that Eq. (\ref{Eq1}) \cite{LoHuJo} is valid 
only for a stationary distribution of the gene-values. However, we have shown that 
the steady-state assumption is only marginally justified for $R=1$, and far from being 
correct for smaller values of the prize-to-fine ratio $R$. 

\begin{figure}
\centerline{\epsfxsize=9cm \epsfbox{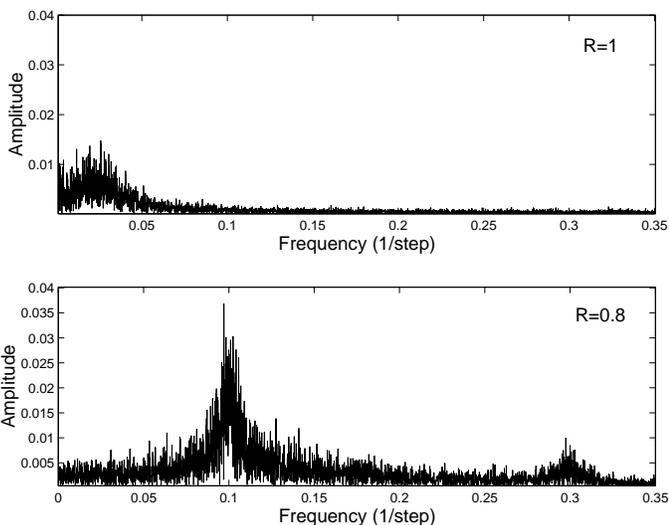}} 
\caption{Fourier transforms of the winning probabilities in the frequency domain for 
$R=1$ (top panel) and $R=0.8$ (bottom panel). The parameters are the same as in Fig. \ref{Fig1}.}
\label{Fig3}
\end{figure}

\begin{figure}
\centerline{\epsfxsize=9cm \epsfbox{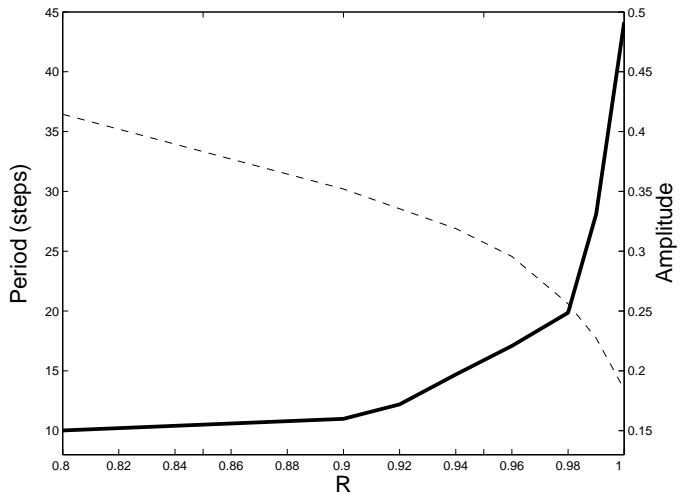}} 
\caption{Dependence of the oscillations period (solid line) and the amplitude (dashed line) 
of the winning probabilities on the value of the 
prize-to-fine ratio $R$. The parameters are the same as in Fig. \ref{Fig1}.}
\label{Fig4}
\end{figure}

To better quantify the temporal oscillations of the winning probabilities, we 
display in Fig. \ref{Fig3} the corresponding Fourier transforms in the
frequency domain. One finds that the transform becomes sharper as the
prize-to-fine ratio decreases (i.e., the 
oscillations are better characterized by a pure, well-defined frequency). 
Figure \ref{Fig4} displays the dependence of the oscillations period (according to the peak of the transform) 
and their amplitude on the prize-to-fine ratio $R$. 
The period of the oscillations decreases with decreasing value of $R$, 
while the amplitude of the oscillations increases with decreasing value of $R$.

We now provide a qualitative explanation for the temporal oscillation which characterize the system. 
Consider for example a situation in which $<$$p$$> <{1 \over 2}$ at a particular instant of time. 
In these circumstances, the winning-probability of an agent with a gene value $p> {1 \over 2}$ is 
larger than {$1 \over 2$} (this is due to the fact that most agents are located in the opposite 
half of the gene-space, and are therefore 
making decisions which are opposite to his decision). 
At the same time, agents with $p< {1\over 2}$ have a small winning-probability, and they are 
therefore losing points on the average. Eventually, the scores of some of these agents fall below $d$, 
in which case they modify their strategy. The new gene-values which are now joining the system lead to 
a global current of gene-values from the $p< {1\over2}$ side of the gene space to the $p> {1 \over2}$ side. 
This increases the value of $<$$p$$>$, and eventually the system will cross from $<$$p$$> < {1 \over2}$ to 
$<$$p$$> > {1 \over2}$. It must be realized that the reaction of the system to this transition is not 
immediate. Agents with $p> {1 \over 2}$ are quite wealthy at this point 
(they had large winning-probabilities in the last few rounds). 
Thus, even tough they start to lose (due to the fact that most of the population is now 
concentrated in their half of the gene space) they do {\it not} modify their gene-values 
immediately. At the same time, 
some of the survived agents with $p< {1 \over 2}$ are quite vulnerable (after losing in the last 
few turns), implying that one wrong choice could drive their score below $d$, forcing them to 
change strategy. In other words, immediately after the crossing 
from $<$$p$$> < {1 \over2}$ to $<$$p$$> > {1 \over2}$, 
agents with $p< {1 \over 2}$ are still more likely to change their strategy. Thus, the average 
gene value continues to increase. Eventually, agents with $p> {1 \over 2}$ (the ones who now 
have poor winning-probabilities) lose enough times and start to modify their strategy. 
This will drive the average gene 
value  back towards $<$$p$$> ={1 \over2}$. This periodic behavior repeats itself again and again, 
producing the temporal oscillations which characterize the system. 

\section{Implications of the temporal oscillations}

The main feature which characterizes the system's behavior is the temporal oscillation of the winning-probability.
In order to capture this effect we consider two types of agents: 
agent A whose winning-probability alternates repeatedly between $1$ and $0$, 
and agent B whose winning-probability, $q$, is {\it constant} in time. 
Agent A represents an extreme agent ($p=0,1$) whose 
winning-probability oscillates in time, while agent B represents a central 
agent ($p={1 \over 2}$) whose winning probability is practically constant in 
time (see Figs. \ref{Fig1} and \ref{Fig2}), and is slightly less than ${1 \over 2}$ \cite{LoHuJo}.

The two types of agents differ in the
standard deviation of their success rate, a fact which dictates a different 
mean life span. 
Consider for example, the simple case of $R=0$ and $d=-1$. 
The mean life span of player A is $1 {1 \over 2}$ rounds (averaging over the two situations: starting the 
game with a victory, or losing in 
the first round of the game). 
The probability of agent B to change his strategy after $n$ rounds 
is $q^{n-1}(1-q)$, and his mean life span is therefore given by $\sum ^{\infty }_{0}nq^{n-1}(1-q)$. 
This equals $\sim 1.98$ for $q=0.495$ (this value of $q$ is taken from the $R=1$ case). 
Thus, agent B has a {\it longer} mean life span. This conclusion is in agreement with the results of the full non-linear 
model (the EMG), in which it was demonstrated that under tough conditions ($R < R_c$) central agents perform 
better than extreme ones (note that this is despite the fact the the average winning probability of a 
central agent is less than that of an extreme agent). On the other hand, for $R=1$ agent A has an infinite life span, while 
agent's B lifespan is finite. Again, this is in agreement with the results 
of the full non-linear model, according to which extreme agents (with large temporal oscillations 
in their winning-probability) live longer than central ones in the $R>R_c$ case.

\begin{figure}
\centerline{\epsfxsize=9cm \epsfbox{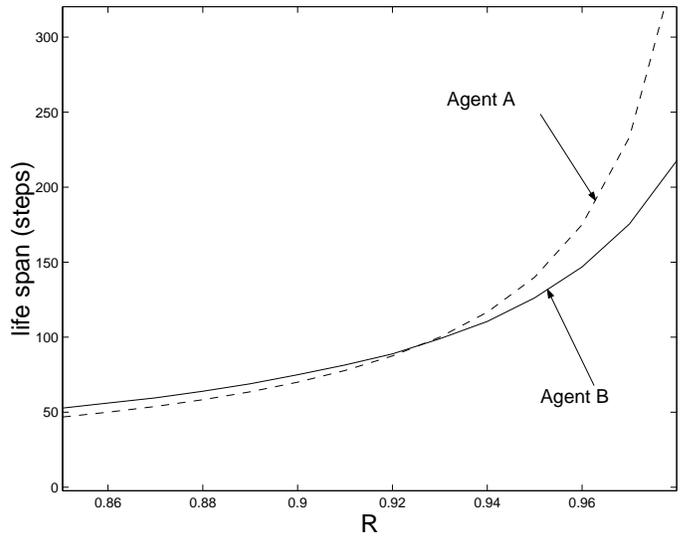}} 
\caption{The life spans of agents A and B as a function of the prize-to-fine ratio $R$. $d=-4$. Agent A wins once 
every second round of the game. Agent B has a constant winning-probability of $0.495$.}
\label{Fig5}
\end{figure}

Figure \ref{Fig5} displays the average lifespan of agents A and B as a function of the prize-to-fine ratio $R$. 
We find that the simplified toy model provides a fairly good qualitative description of the 
complex system. In particular, in the $R=1$ case agent A (the extreme one) 
performs better (with a longer mean lifespan) than agent B (the central one), in agreement with the 
fact that the population tends to self-segregate into opposing groups characterized by extreme 
behavior \cite{John1}. On the other hand, 
for $R<R_c$ agent A performs worse, in agreement with the finding \cite{HodNak} that 
in times of depression the population tends to cluster around $p={1 \over 2}$. 

The simplified model can explain another interesting feature of the full EMG: 
it was found in \cite{HodNak} that the relative 
concentration [$P(0):P({1 \over 2})$] 
of agents around $p=0$ (and $p=1$) in the $R=1$ ($R>R_c$) case is larger than the 
relative concentration [$P({1 \over 2}):P(0)$] 
of agents around $p={1 \over 2}$ in the $R=0.971$ ($R<R_c$) case 
(see Fig. 1 of \cite{HodNak}). 
This result can be explained by the fact that the lifespan difference between 
the various agents is larger in the 
$R=1$ case as compared with the $R<R_c$ case (see Fig. \ref{Fig5}).

It should be realized that in order to have a long average lifespan in the $R>R_c$ case, 
it is best {\it not} to take unnecessary risks. 
An agent who plays with a constant (i.e., time independent) 
winning probability (agent B) 
takes the risk of losing more times than he wins (and this may derive his score below $d$). 
The average life span of agent B is therefore 
shorter than the corresponding average life span of agent A who wins and loses exactly 
the same number of times. 

On the other hand, in the $R<R_c$ case, agents {\it must} take risks in order 
to survive. An individual agent cannot afford himself 
to win and lose the same number of times (since the fine is larger than the prize). 
In order to survive under harsh conditions an 
agent {\it must} win more times than he lose. Thus, in such conditions ($R<R_C$) agent B has 
a longer average life span as compared with 
agent A (Playing with a constant winning probability is 
the best strategy to achieve more winnings than loses.) 

\section{Summary and discussion}

In summary, we have considered a semianalytical model of the 
evolutionary minority game with an arbitrary value of the prize-to-fine ratio $R$. The 
main results and their implications are as follows:

(1) The winning-probabilities of the agents display temporal-oscillations. 
The smaller the value of the prize-to-fine ratio $R$, the farther the system is 
from a steady-state distribution (the larger is the amplitude of the oscillations). 
Extreme agents (ones with $p=0,1$) have larger 
oscillations in their winning probability as compared with central ($p={1 \over 2}$) agents. 
Thus, extreme agents are 
sensitive to the global gene-distribution of the population (their winning-probabilities display large 
temporal oscillations), while central agents have 
an almost constant (time-independent) winning probability ($\sim {1 \over 2}$).

(2) In the $R>R_c$ case the population tends to {\it self-segregate} into {\it opposing} 
groups. The winning probabilities of these two groups oscillate in time in such a way that each 
group wins and lose approximately the same number of times. The efficiency of the system is therefore
maximized due to the fact that at each round of the game one of the groups (containing approximately 
{\it half} of the population) wins. Thus, by self-segregation into two opposing groups, the agents 
cooperate indirectly to achieve an optimum utilization of their resources.

On the other hand, in the $R<R_c$ case an individual agent cannot afford himself 
to win and lose the same number of times. In order to survive under harsh conditions ($R<R_c$) an 
agent {\it must} win more times than he lose. Thus, in a tough environment agents cannot 
cooperate (not even indirectly) by self-segregating into two opposing groups. 
Rather, they tend to {\it cluster} around 
$p={1 \over 2}$. Playing with a constant (i.e., time-independent) winning probability ($\sim {1 \over 2}$) provides 
an individual agent with the best chance to win more times than he loses 
[an extreme agent on the other hand (with large oscillations in his winning probability) wins
and loses approximately the same number of times]. Note that while 
playing with a constant winning probability is the only way to survive in a tough environment (the only way to win 
more times than losing), it is also 
the riskiest strategy: such an agent takes the risk of losing more times than he wins. 

The clustering phenomena creates a situation in which the population as a whole is {\it not} organized. 
Due to statistical fluctuations, the average number of winners at each round of the game is less than half of the 
population, implying a low efficiency of the system as a whole. 

\bigskip
\noindent
{\bf ACKNOWLEDGMENTS}
\bigskip

The research of SH was supported by grant 159/99-3 from the Israel 
Science Foundation and by the Dr. Robert G. Picard fund in 
Physics. The research of EN was supported by the Horwitz foundation.


\begin{thebibliography}{99}
\bibitem{note1} The model has a symmetry under the transformation $p \to 1-p$.
\end{thebibliography}
\end{document}